\documentclass[journal]{IEEEtran}
\usepackage{comment,cite}
\usepackage{booktabs} 
\usepackage{graphicx,adjustbox,float}
\usepackage{multirow,tikz}
\usepackage[colorinlistoftodos]{todonotes}
\usepackage{xcolor}
\newcommand\ddfrac[2]{\frac{\displaystyle #1}{\displaystyle #2}}
\usepackage{color, colortbl}

\usepackage{algorithm,algorithmic}
\usepackage{amsbsy,amsmath,amssymb,epsfig,bbm,mathrsfs,fancyhdr,fancyvrb,subfigure,url}
\usepackage{pgfplots}

\usetikzlibrary{patterns}

\begin{document}
\title{CUDA-Accelerated Application Scheduling in Vehicular Clouds Under Advanced Multichannel Operations in WAVE}

\author{Yassine~Maalej,~\IEEEmembership{Student~Member,~IEEE,}
        and~Elyes~Balti,~\IEEEmembership{Student~Member,~IEEE} }

\maketitle

\begin{abstract}
This paper presents a novel Advanced Activity-Aware (AAA) scheme to optimize and improve Multi-Channel Operations based on the IEEE 1609.4 standard in wireless access vehicular environments (WAVE). The proposed scheme relies on the awareness of the vehicular safety load to dynamically find an optimal setup for switching between service channel intervals (SCHI) and control channel intervals (CCHI). SCHI are chosen for non-critical applications (e.g. infotainment), while CCHI are utilized for critical applications (e.g. safety-related). We maximize the channel utilization without sacrificing other application requirements such as latency and bandwidth. Our scheme is implemented and evaluated network simulator-3 (NS3). We guarantee the default Synchronization Interval (SI), like implemented by the standard in vehicular ad hoc networks (VANETs), when tested on real-time simulations of vehicular cloud (VC) load and VANET setups using NS3. We also evaluate a Markov Decision Process (MDP) based scheme and a fast greedy heuristic to optimize the problem of vehicular task placement with both IEEE 1609.4 and an opportunistic V2I version of the proposed AAA scheme. Our solution offers the reward of the VC by taking into account the overall utilization of the  distributed virtualized VCs resources and vehicular bag-of-tasks (BOTs) placements both sequentially and in parallel. We present the simulation metrics proving that our proposed solution significantly improve the throughput and decreases the average delay of uploaded packets used for non-safety applications, while maintaining reliable communication (via CCHI) for safety-related applications similar to the IEEE 1609.4 standard.
\end{abstract}

\begin{IEEEkeywords}
vehicular cloud (VC), CUDA, dedicated short range communication (DSRC), wireless access vehicular environments (WAVE), vehicle to infrastructure (V2I), vehicle to vehicle (V2V), Markov decision process (MDP), IEEE 1609.4, network simulator-3 (NS3), service channel (SCH), Control channel (CCH).
\end{IEEEkeywords}

\IEEEpeerreviewmaketitle

\section{Introduction}
\IEEEPARstart{V}{ehicular} clouds (VCs), establishing synergies between cloud computing and vehicular ad-hoc networks (VANETs), are the key corner stone for building safer roads for smart cities, advanced management for transportation, and on over the air connected and streamed infotainment services \cite{RUBIN2019125}.
The major utilization of VCs is to ingest data and applications that are continuously requested and used on VANETs across its multiple constituting vehicles. The virtualized computational resources in on board units (OBUs) of vehicles and in road-side units (RSUs) represent the key computing parts to host IP based non-safety infotainment services. The reliability of the vehicular network and the VC throughput and delay are of paramount importance due to the major technologies integration they need to handle, including the increase of data offloaded from the cloud, rapidly changing network topology, and vehicle/pedestrian safety. The dedicated short range communication (DSRC) \cite{DesignOfGHzDSRCBasedVehicularSafetyCommunication} standard is key to effective vehicular communication technology \cite{FederalCommunicationsCommissionFCCReportAndOrderStd}. In the USA, DSRC spans over a dedicated 75 MHz spectrum band around 5.9 GHz \cite{SafetyCommunicationsInTheUnitedStates}. The used bandwidth is partitioned into seven different channels of 10 MHz each. The channels contain one Control Channel (178) to serve event-driven and safety related applications, and six Service Channels (172,174,176,180,182,184) to serve packets of cloud-based non-safety applications in vehicular networks. Time critical safety applications are based on vehicle to vehicle (V2V) communication and delay-tolerant non-safety applications are based on vehicle to infrastructure (V2I) communication, and can cover addition V2V safety messages while the infrastructure connectivity near it limits. Metrics such as latency, coverage as well as security can be improved using different techniques. Latency, security and data rate for example can be enhanced using millimeter wave (mmWave) communications \cite{balti2020mmwaves,neji1,balti2019tractable,neji2} or high directional antennas design \cite{balti2019sub6} to compensate for the pathloss and increase the received power. Moreover, vehicular applications such as platooning, collaborative perception, video games sharing and autonomous driving require high coverage and spectral efficiency requirements for big data processing. These metrics can be enhanced using relaying assisted cooperative communications such as half-duplex \cite{c1,c2,c3,j1,j2,balti2020joint,j3,eT,asym}, full-duplex \cite{balti2020zeroforcing,balti2020adaptive}, and intelligent reflecting surfaces \cite{mensi2020pls,balti2020stochastic,mensi2020physical}. Consequently, the realistic and successful deployment of VANETs  would require a certain level of safety messages reliability as well as to take maximize the utilization of the abundant compute capabilities in VCs to ingest data and service with a delay and jitter being as low as possible while the virtual machine migration (VMM) occurs, all while maintaining the resource for the coalition forming the cloud as presented in \cite{Journal_sensor_fusion}.
\par The developed claims and results in this paper are summarized as follows:
\begin{itemize}
    \item Improve the VCs connectivity by using at maximum capacity the service channel (SCH) and lowering the end-to-end delays of V2I packets.    
    \item Implement and test a new approach for DSRC channel inactivity detection and automatically achieve a  dynamic switching between different intervals based on the average channel workload.    
    \item Improving the the number of the VMs drops during the migration by achieving a high throughput of VC and enhancing the end-to-end delay of exchanged packets between RSUs and the vehicles' OBUs.  
    \item Develop and achieve an optimal vehicular task scheduler scheme as a Markov decision process (MDP) and comparison with a fast greedy algorithm for scheduling vehicular tasks in a vehicular cloud computing (VCC) system containing numerous computing VCs with opportunistically available V2I communication.  
    \item We define an accelerated version of the iteration value algorithm to determine optimal scheduling epochs of the MDP formulation using CUDA GPGPU parallel compute.
\end{itemize}
\par The rest of this paper is organized as follows: in section \ref{sec2:AdvancedActivityAwareMultiChannelOperation}, we develop the proposed advanced activity-aware (AAA) multi-channel operations scheme for maximization of SCH utilization. In Section \ref{WorkloadModelingVehicularCloudModeling}, we show the architecture of the VCs and workload modeling of the VC tasks. In Section \ref{FastGreedyHeuristicsForSchedulingInVehicularCloudCompution}, the fast greedy scheduling algorithm is developed.
In section \ref{SequentialAndCUDAacceleratedMDPBasedscheme}, both of the sequential and the CUDA accelerated schemes for vehicular task scheduling are discussed. In Section \ref{sec3:SimulationAndComparison}, we report the simulation results and present an evaluation of the enhancements of applications' metrics in VANETs under the proposed AAA scheme.  In Section \ref{NumericalResultsAndAnalysismdpgreedySIMULATION}, we discuss the numerical results, performance analysis and evaluation of the developed schemes in terms of VCC resource utilization and reward after scheduling decisions. Finally, we conclude the paper by summarizing the findings of this article in Section \ref{SectionConclusion}.
\subsection{WAVE 802.11p MAC }
\begin{figure}[H]
\begin{center}
\includegraphics[width=\linewidth]{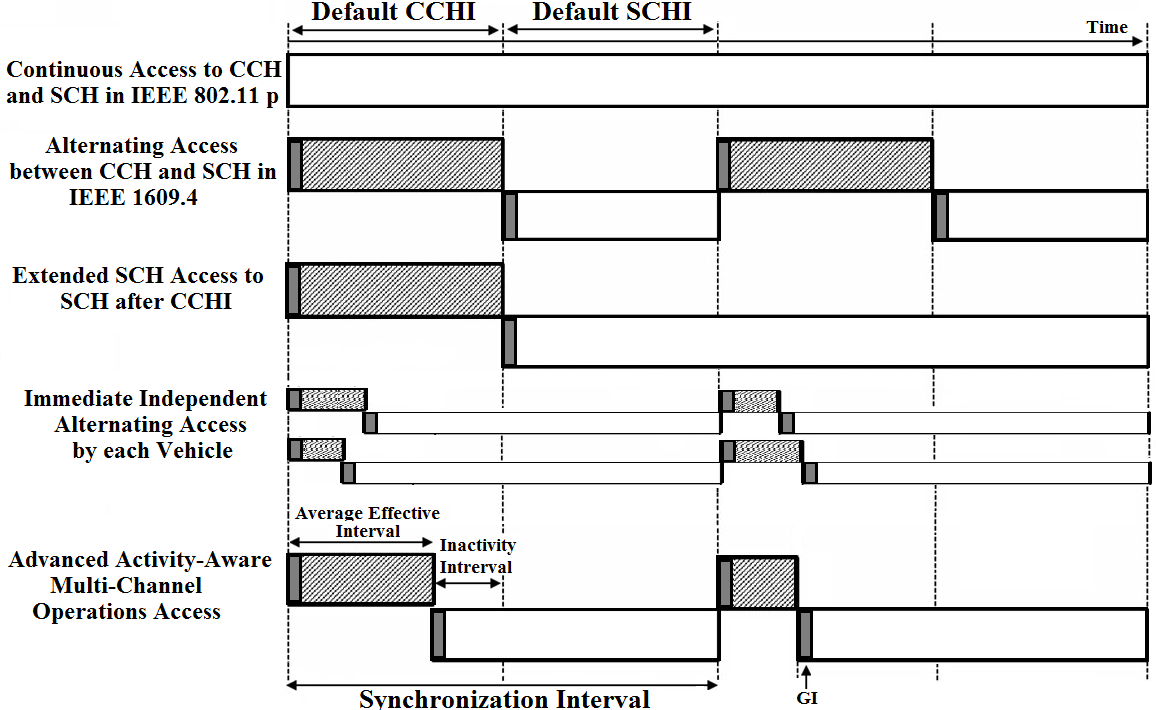}
\caption{WAVE Different Channel Access Mechanisms.}
\label{WaveDifferentChannelAccessMechanisms}
\end{center}
\end{figure}

\begin{figure}[!t]
\begin{center}
\includegraphics[width=8cm,height=8cm]{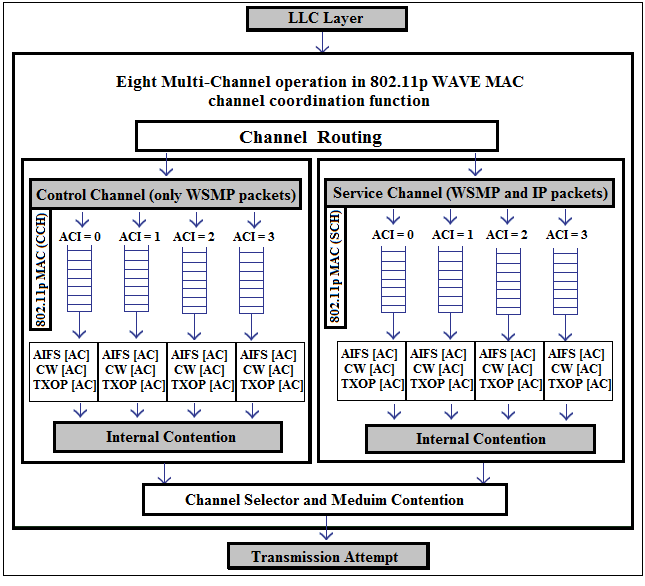}
\caption{IEEE 1609.4 MAC With Multi-Channel Operations.}
\label{IEEE16094MACithMultiChannelOperation}
\end{center}
\end{figure}

Most of the Medium Access Control (MAC) protocols developed in the literature and used in IEEE 802.11p are employed in earlier standards of Wireless LANs. The medium access is very challenging in VANETs because of rapidly changing network topologies due to mobility. Frequency Division Multiple Access (FDMA) \cite{ANewApproachToExploitMultipleFrequenciesInDSRC}, Time Division Multiple Access (TDMA) \cite{ANovelCentralizedTDMABasedSchedulingProtocolForVehicularNetworks} and Code Division Multiple Access (CDMA) \cite{AStudyOfInterVehicleCommunicationSchemeAllocatingPNCodesToTheLocationOnTheRoad} are very complex to implement because of the permanent need of coordination to associate frequency channels, time slots or codes with vehicles. The IEEE 802.11p MAC protocol uses Carrier Sense Multiple Access with Collision Avoidance (CSMA/CA) protocol \cite{IEEE80211pPerformanceEvaluationAndProtocolEnhancement} that is highly considered the most suitable for VANETs. IEEE 802.11p offers to transmit packets on one channel and employs a continuous mechanism \cite{ImpactOfIEEE16094ChannelSwitchingOnTheIEEE80211pbeaconingperformance} for DSRC channel access to send both non-safety and safety packets, as presented in Fig. \ref{WaveDifferentChannelAccessMechanisms}.
WAVE recommends that, for a fixed SI equal to 100ms, every vehicle needs to be tuned to the CCH in order to exchange their safety related messages with neighboring vehicles as detailed in \cite{OptimizingtheControlChannelIntervaloftheDSRCforVehicularSafetyApplications} and in \cite{Conf_Sensor_Fusion}.
WAVE differs between various Quality of Service (QoS) levels by using specific fields in the packets that are sent over CCH or SCH. It supports contention-based Enhanced Distributed Channel Access (EDCA) mechanism \cite{ImpactOfIEEE16094ChannelSwitchingOnTheIEEE80211pbeaconingperformance}. EDCA guarantees four access levels and classes using different transmission queues, as detailed in Fig. \ref{IEEE16094MACithMultiChannelOperation}, as well as variable Arbitration Inter-Frame Spacing (AIFS) periods, minimum and maximum CWs for traffic priority. To guarantee channel switching and coordination mechanisms, the IEEE 1609.4 standard for Multi-Channel Operations \cite{IEEE16094DSRCmultichanneloperationsanditsimplicationsonvehiclesafetycommunications} is developed on top of IEEE 802.11p MAC as Fig. \ref{IEEE16094MACithMultiChannelOperation} presents.
\subsection{IEEE 1609.4 Standard for Multi-Channel Operations}
The IEEE 1609.4 standar offers many wireless radio accesses for WAVE by coordinating managing the DSRC channel switching as specified in the IEEE 802.11p MAC and Physical Layer (PHY) layers. Fig. \ref{WaveDifferentChannelAccessMechanisms} details the usage of the intervals between channels with divided access times based on alternatively switching between CCH and SCH intervals for safety packets and non-safety messages, respectively. In the scenario of VANETs with few vehicles, the static alternating scheme defined in IEEE 1609.4 does not satisfy a best effort channel usage. It mandates that vehicles should tune to CCH (resp SCH) during CCHI (resp SCHI) independently of the queued messages and the required rates of messages generation. Networks with low density have a significant inactivity for safety channels during the timer specified in IEEE 1609.4. It is considered as a big incentive to achieve a dynamic fully utilization timers to grant access for the various types of channels based on dynamic settings of the VANETs' safety and non-safety packets transmission, as detailed in \cite{7841571}. Our advanced and optimal access improves the throughput, allows the VCs to manage and optimize the task placement and allows higher VM migration from vehicles before leaving the network with minimal content loss as detailed in \cite{8746657}.
\section{Advanced Activity-Aware (AAA) Multi-Channel Operations}\label{sec2:AdvancedActivityAwareMultiChannelOperation}

in Table. \ref{SymbolsAndNotations}, we present the various setting and variables to model the efficiency of the AAA scheme. \ref{SymbolsAndNotations}. Our AAA Multi-Channel Operations solution dynamically changes the MAC timers after calculating the effective usage of CCH to broadcast safety beacons. Non static vehicles in the range of each RSU will have various changing distances between each others, and the network will face variable end-to-end delays of the BSMs sent in CCHI. In addition, vehicles entering the network and the VC will have to send non-safety packets and keep broadcasting safety beacons. The global awareness of the network density is used to approximate the average end-to-end delay and rate and numbers of BSMs to dynamically reduce the inactivity time over CCH accesses. First, we begin by learning the characteristics and statics of V2V over CCH communication during BSMs broadcasting. Second, each RSU captures the details of the vehicles under its coverage,  then it executes the AAA algorithm. 
The rates of the safety messages \textit{$RG_{BSM}$} helps the calculation of the overall number queued BSMs in every SI by using Eq.\ref{NGBSMPERVEHICLEPERSI} and  it represents ready in non-sent packets at the beginning of every Guard Interval (GI) when the timer roll over for CCHI.
\begin{equation} \label{NGBSMPERVEHICLEPERSI}
NG_{BSM}=RG_{BSM}*GI
\end{equation}
The repeated CCH access during the CCHI, the successfully acknowledged BSMs in vehicle's OBU is determined by Eq.\ref{BSMSentPerSI}  as the overall difference between the created and queued BSMs packets \textit{$Qv_{h}$} whenever the switching of channel happens. 
\begin{equation} \label{BSMSentPerSI}
Sv_{h}= NG_{BSM}-Qv_{h}
\end{equation}
The end-to-end delay of the acknowledged BSMs is determined by \textit{$R_{vhBSM}$}, for each vehicle $vh$ by solving the Eq. \ref{DelayReceivedBSMbyVehicle}. Let \textit{$t_{Rijh}$} and \textit{$t_{Sijh}$} be respectively the time of \textit{$i^th$} BSM acknowledged by vehicle $vh$ and initiated by vehicle $vj$. 
\begin{equation} \label{DelayReceivedBSMbyVehicle}
Dv_{h}=\ddfrac{\sum_{i=1, i\neq h}^{N_{V}} \sum_{j=0}^{N_{ih}} t_{Rijh} - t_{Sijh}} {R_{vhBSM}}
\end{equation}  
The lack of activity during current switching scheme is measured by difference between the static CCHI and the average effectively used CCHI, is determined by Eq. \ref{AverageEffectiveCCHUtilization}, where \textit{$D_{V2V}$},given by Eq. \ref{AverageDelayV2V}, represents the average delay of BSM in V2V communication and \textit{$S_{V2V}$}, given by Eq. \ref{SentV2VBSMS}, is the number of exchanged BSMs in the network.
\begin{equation} \label{AverageEffectiveCCHUtilization}
U_{CCH}= S_{V2V}*D_{V2V} 
\end{equation}  
\begin{equation} \label{AverageDelayV2V}
D_{V2V}= \frac{ D_{V2V} + \sum_{h=1}^{N_V} D_{vh}  } \\
\end{equation}  
\begin{equation} \label{SentV2VBSMS}
S_{V2V}= \frac{S_{V2V} + \sum_{v_{h}=1}^{N_{V}} Sv_{h}}{1+N_{V}} \\
\end{equation}  
The average increase of successfully exchanged packets from vehicles to RSU is calculated by Eq. \ref{IncreasedVCPackets}, during the time of extension of  SCHI that is formulated in Eq. \ref{UpdateServiceChannelInterval} based on the calculated inactivity interval given by Eq. \ref{InactivityInterval} as the difference between the default interval and the effective utilization interval .
\begin{equation} \label{InactivityInterval}
I_{CCH}= CCHI - U_{CCH}  
\end{equation}  
\begin{equation} \label{UpdateServiceChannelInterval}
ESCHI = DSCHI + ICCH  
\end{equation}  
\begin{equation} \label{IncreasedVCPackets}
S_{VC}= \sum_{h=1}^{N_{V}} NG_{VC}-Qv_{{i}_{VC}} 
\end{equation}  

\section{Vehicular Task Workload Modeling and Vehicular Cloud Computing Model }\label{WorkloadModelingVehicularCloudModeling}
\subsection{Vehicular Task Workload Modeling}
We use a set of \textit{n} Bag-of-Tasks (BOTs), as presented in Fig. 14, every BOT has a number of independent tasks that can be executed in any order or time. Every BOT requirements in delay, throughput and number of VMs have to be satisfied by the VC capacities (e.g. delay and throughput). For the case of VCC, we do not consider any temporal bursts of the tasks to arrive and we assume the readiness of the BOTs before any scheduling decision is made. 
\begin{table}[!t]
\begin{center}
\caption{VARIABLES NOTATIONS}\label{SymbolsAndNotations}
\begin{tabular}{ll}
\toprule 
Variable & Description  \\
\midrule
\rowcolor{black!20} 
$N_{V}$         & Network Vehicles Density  \\
$ST$            & Duration of Simulation  \\
\rowcolor{black!20} 
$SI$            & Static SI  \\
$N_{Sync}$      & Count of SIs in Simulation period    \\
\rowcolor{black!20} 
$SI_{i}$       & Current SI    \\
$I_{CCH}$       & II for CCHI    \\
\rowcolor{black!20} 
$U_{CCH}$       & Averaged CCH Utilization    \\
$RG_{BSM}$      & BSM packet generation rate        \\
\rowcolor{black!20} 
$RG_{VC}$       & VC non-safety application packet generation rate \\
$NG_{BSM}$      & Total BSM messages generated by vehicle  \\
\rowcolor{black!20} 
$NG_{VC}$       & VC apps packets number per vehicle  \\
$S_{vi}$  & Total BSM generation by vehicle $vi$  \\
\rowcolor{black!20} 
$D_{vi}$  & Average end-to-end delay of received BSMs by $vi$   \\
$Q_{vi}$  & Number of queued packets BSM of $vi$ \\
\rowcolor{black!20} 
$Qvi_{VC}$   & Number of queued packets VC  of $vi$ \\
$Rv_{{i}_{BSM}}$  & Number of BSM received of vehicle $vi$  \\
\rowcolor{black!20} 
$t_{Sijh}$  & Time $j^{th}$ BSM is sent by $vi$ and received by $vh$ \\
$N_{ih}$    & Total number of BSMs sent by $vi$ and received by $vh$   \\
\rowcolor{black!20} 
$tSVC_{ijRSU}$ & Time packet j is sent from $vi$ to the RSU  \\
$tRVC_{ijRSU}$ & Time packet j sent from $vi$ is received by RSU \\
\rowcolor{black!20} 
$NiRSU$        & Packets VC sent by $vi$ and received by RSU \\
$S_{V2V}$       & BSMs Average number for all vehicles \\
\rowcolor{black!20} 
$S_{VC}$        &  VC packets average vehicles to the RSU \\
$D_{V2V}$       & End-to-end average delay of BSMs \\
\rowcolor{black!20} 
$D_{VC}$        & End-to-end average delay for VC packets \\
$t_{Rijh}$  & Time $j^{th}$ BSM sent by $vi$ is received by $vh$ \\ 
\rowcolor{black!20} 
$DCCHI$         & Default CCHI  \\
$DSCHI$         & Default SCHI  \\
\bottomrule
\end{tabular}
\end{center}
\end{table}
Algorithm l presents more details on finding the average end-to-end delay between exchanged BSMs between every SI transition, as well as details about inactivity and increased SCHI in order to keep the default SI equal to 100 ms as predefined by the standard. 
\begin{algorithm}[t]
\caption{AAA Multi-Channel Steps}
\label{algorithone}
\begin{algorithmic}[1]
 \STATE \textbf{Result:} Updated CCHI and SCHI 
  \STATE $N_{Sync}= \frac{ST}{SI} , SI_{i}=1$ 
  \WHILE{$SI_{i} < N_{Sync}$}
     \IF{Current Time != UTC second}
       \STATE         $S_{V2V}= \frac{S_{V2V} + \sum_{v_{h}=1}^{N_{V}} Sv_{h}}{1+N_{V}} $
       \STATE        $D_{V2V}= \frac{ D_{V2V} + \sum_{h=1}^{N_V} D_{vh}  }   
        { 1+ \sum_{v_{h}=1}^{N_{V}}  Rv_{{h}_{BSM}} } $
     \ELSE
       \STATE Average Effective CCH Utilization
       \STATE $ U_{CCH}= S_{V2V}*D_{V2V} $
       \STATE Inactivity Interval in current CCHI
       \STATE $ I_{CCH}= CCHI - U_{CCH} $
          \IF{$U_{CCH} < D_{CCHI}$}
            \STATE Update MAC Interval Timers
            \STATE $ CCHI= U_{CCH} $
            \STATE $ SCHI= SCHI + I_{CCH} $
            \STATE The $NG_{VC}$ are ready at the new SCHI
            \STATE $NG_{VC}=RG_{VC}*SI $ 
            \STATE Increased $S_{VC}$ in new longer SCHI
            \STATE $S_{VC}= S_{VC} + \sum_{h=1}^{N_{V}} NG_{VC}-Qv_{{i}_{VC}}$
            \STATE Reduced $D_{VC}$ in the new longer SCHI
            \STATE             $D_{VC}= \frac{ D_{VC} + \sum_{i=1}^{N_V} \sum_{j=0}^{NiRSU}  tRVC_{ijRSU}-tSVC_{ijRSU}}   
            { 1+ \sum_{i=1}^{N_V} \sum_{j=0}^{NiRSU}  tRVC_{ijRSU}-tSVC_{ijRSU}} $
         \ELSE
           \STATE The number of vehicles is high and CCHI is not sufficient
           \STATE $ CCHI = D_{CCHI} $ 
           \STATE $ SCHI = D_{SCHI} $ 
         \ENDIF
        \STATE $S_{V2V}=0$
        \STATE $D_{V2V}=0$
        \STATE $SI_{i}$= $SI_{i}$ + 1 
     \ENDIF
  \ENDWHILE
\end{algorithmic}
\end{algorithm}

\subsection{Modeling of VCC }\label{VehicularCloudModeling}
The AAA developed scheme and the default IEEE 1609.4 are compared with the simulation results shown in Fig. \ref{VcThroughputOfNonSafetyApplications} as throughput under various VCs formulations with RSUs computes. To simplify, we assume that every vehicle contributes in the VC creation with one VM virtualized as its OBU micro compute. The general VCC cluster is created based on various VCs and changes from one vehicular network to another. Each group of vehicles with every RSU range form one VC, as detailed in Fig. \ref{VCWithRSUgatewaySimulationInNS3},  with various OBUs used as VMs, density, VM throughput and delay. Our developed AAA model determines the end-to-end average delay as well as the total count BSMs successfully exchanged. It helps to evaluate dynamically if the considered CCHI can be squeezed or not to increase SCHI as presented in Fig. \ref{WaveDifferentChannelAccessMechanisms}. 
We notice significant improvement in ever VC metric by using the AAA over the scheme in IEEE 1609.4. The problem of scheduling the tasks of BOTs into the available resources in VCs is achieved while taking into account the uniqueness of each task and maximizing VMs utilization.

\section{Fast Greedy Heuristics for scheduling vehicular tasks in Vehicular Cloud Computing} \label{FastGreedyHeuristicsForSchedulingInVehicularCloudCompution}
Our heuristic fast greedy solution published in \cite{ApplicationOfGreedyAlgorithmsToVirtualMachineDistributionAcrossDataCenters} detects the sub-optimal result for the problem of the VM scheduling and allocation. The quick scheduling greedy solution does not evaluate previous choices made after each scheduling decision epochs. Consequently, it will find non-optimal decisions because one policy scheduling can give the highest instant reward among various scheduling decisions, but it may reduce the overall reward of the following scheduling decisions. Our solution is applied to the total VCC that is the group of the various VCs forming micro data centers for the compute as detailed in the model above. Based on the previous works in literature, there are no  direct quick decision-making solution to optimally schedule tasks in VCs. Consequently, the optimal set of scheduling policies can vary based on the the distribution and availability of VMs deployed in VCs and based on the vehicular non-safety services requirements in every BOT. The greedy-assignment of vehicular application tasks and processes to the available VMs in every VC that is developed in \cite{GreedySchedulingOfTasksWithTimeConstraintsForEnergyEfficientCloudComputingDataCenters} sorts the available VMs in every VC by sorting the performance metrics and schedules every task to the most suitable hosting VMs. Then, repeat the allocation process of the remaining tasks in BOTs and stops when all the VMs in the VCs are used. Our developed quick greedy heuristic solution is presented by the following Algorithm \ref{AlgorithmGreedyScheme}.

\begin{algorithm}[t]
\caption{Fast Greedy allocating efficient VM first}
    \label{AlgorithmGreedyScheme}
\begin{algorithmic}[1]
\STATE \textit{VCC} $\gets$ \text{Set of vehicular clouds forming the VC}
    \STATE \textit{$S_{BOT}$} $\gets$ \text{Set of different BOT} 
    \STATE \textit{$T_{i}$} $\gets$ \text{Set of tasks in the $i^{th}$ BOT} 
    \STATE \textit{most efficient VM (mevm)} $\gets$ NULL
    \FORALL{ bot$_{i}$ $\in$ S$_{BOT}$}
        \FORALL{ T$_{ij}$ $\in$ bot$_{i}$  }
            \STATE isScheduled=Flase
            \FORALL{vc$_{i}$ $\in$ VCC} 
                \IF{isScheduled=True}
                    \STATE Break
                \ENDIF
                    \FORALL{$VM_{i}$ $\in$ vc$_{i}$} 
                        \IF{$VM_{i}$ is available and meets the requirements of the task $T_{ij}$}
                        \STATE mevm = $VM_{i}$, isScheduled=True, $T_{ij}$ $\notin  bot_{i}$
                        \ENDIF
                    \ENDFOR
            \ENDFOR
            \IF{isScheduled=Flase}
                \STATE Task T$_{ij}$ is placed in paid cloud
                \STATE T$_{ij}$ is removed from bot$_{i}$ 
            \ENDIF
        \ENDFOR
    \ENDFOR
\end{algorithmic}
\end{algorithm}


\section{CUDA-accelerated and Sequential MDP-Based solution for processes and tasks scheduling in VC Computing}\label{SequentialAndCUDAacceleratedMDPBasedscheme}
\subsection{MDP-Based Scheme Solution Model Formulation}
Abstractly, the MDP is composed of a state space S of VCC, a group of actions A of possible scheduling epochs, function for immediate reward r (s,a,s') of the entire set of VCs and the probabilistic transition function P (s'$|$s,a) from one state \textit{s} to another state \textit{s'} after taking the scheduling action \textit{a}.
\subsection{System States}
As described above, the focus is to schedule many tasks in the BOTs that are requiring various VMs from VCs of the entire VCC system. Although we want to achieve an efficient scheduling of tasks, there is a trade-off to ensure between that we minimize the cost paid VMs to host the tasks and we maximizing the complete total VCC reward. The set of state S of the MDP problem represents the set of geographically distributed VCs and the corresponding VMs in OBUs, the process and tasks of the services, and the services that still need to be hosted. Namely, we denote the set of states as follows: 
\begin{equation} \label{SystemStates}
    S=  \{ s|s = (n1,n2,...,n_{TT}, M,e) \} 
\end{equation}
where $n_{i}$ represents the total number of vehicular tasks allocated to the $i^{th}$ VC, the symbol M is the available set of VMs in the current VC and the symbole $e$ is request to place a task. 
\subsection{Actions to Schedule}
The MDP model to schedule tasks have numerous random possible actions to schedule \textit{a} from the actions set:     
\begin{equation} \label{Actions}
    A = \{-1,1\}
\end{equation}
With the every incoming task scheduling request, the VCC selects an action from $A$ that has to be executed. Scheduling in paid TCC has a reducing value of the reward by -1. It is caused by the the throughput limitation, communication delay or lack of VMs. Whereas we set 1 as a value to characterise the scenario when the picked VC guarantees the requirements of the minimum delay and VM of the task.  
\subsection{Transition function between states after scheduling epochs}
Our model is characterized by a probabilistic transition function P (s'$|$s,a) $\longrightarrow$ [0,1] to determine the probability to transition from one state \textit{s} to a next state \textit{s'} in the VCC after executing a scheduling action \textit{a}. With all the tasks scheduling requests, there will be check if any hosting VC guarantees the average packet delay for V2I is in accordance with the requirement of the tasks and whether or not it has free VMs with the required throughput by the process of the vehicular application.  
After a task placement request is submitted, the total number of VMs occupied by other tasks is given by Eq.\ref{NumberOfOccupiedVMs} no matter of action taken by the VCC.
\begin{equation} \label{NumberOfOccupiedVMs}
N_{v}= \sum_{i=1}^{N_{c}} \sum_{j=1}^{N_{a}} u_{ij}+v_{j} 
\end{equation}
After scheduling request is served of a task \textit{j}, that requires $n_{j}$ VMs, is placed and executed on \textit{i$^{th}VC$}, then the updated number VMs that are not free in the VC is 
\begin{equation} \label{NumberOfOCuupiedVMsPerVC} 
T_{VMi}= \sum_{i=1}^{R_{c}}u_{i} + n_{j} 
\end{equation}  
\subsection{Rewards}
After the scheduling action \textit{a} is taken, the immediate reward of the system in the actual VCC state \textit{s} is defined by: \\
\begin{equation} \label{Actions}
    r(s,a,s') = k(s,a,s') - g(s,a,s')
\end{equation}
\textit{k(s,a,s')} denotes the instant VCC reward just after executing the scheduling action \textit{a} from the previous state \textit{s} for a scheduling event \textit{e} of scheduling tasks is complete. It characterises the immediate reward of the hosting VC of the task. 
\textit{g(s,a,s')} represents the immediate cost after executing the task placement request of the scheduling epoch. 
After a scheduling request is admitted to the VC, then the instant reward is determined by Eq. \ref{InstantReward} and consequently the cost of scheduling is expressed by Eq. \ref{InstantCost}.
\begin{equation} \label{InstantReward}
    k(s,a,s') =  \beta_{vc}* n_{j}
\end{equation}
\begin{equation} \label{InstantCost}
    g(s,a,s') = \beta_{tc}*n_{j} + \gamma_{vc}*\mu_{i} 
\end{equation}
$\mu_{i}$ represents the free underutilized VMs of the $i^{th}$ VC. $\beta_{vc}$, $\beta_{tc}$ and $\gamma_{vc}$ are the immediate reward of VM per VC, cost of scheduling of VM per TCC and the penalty of underutilized free VM in VCs, respectively. The VCC system is considered as a loosing solution without any reward when it reached all states with VMs of each VC are completely full and used. Consequently, the scheduling decisions of the remaining tasks will create a negative reward since they will require cost to be hosted in paid VMs. 
\subsection{Total Reward of the Solution}
The main objective of the MDP-based task scheduling in VCC is to optimize the task placement of various BOTs over distributed VCs that increases the function of the total cloud reward. The maximum expected long-term cumulative reward of the VCC could be expressed as:
\begin{equation} \label{totalReward} \begin{split}
R &= \max_{s \in S} [\sum_{s' \in S, a \in A}{ r(s,a,s')}] \\
  &= \max_{s \in S} [\sum_{s' \in S, a \in A}{ k(s,a,s') - g(s,a,s')}] \\
  &= \max_{s \in S} [\sum_{s' \in S, a \in A}{ (1 + \beta_{vc})*n_{j} - (\beta_{tc}*n_{j} + \gamma_{vc}*VC_{ij})}]\\
\end{split} \end{equation}
\subsection{MDP Optimal Solution Scheduling }
The best solution of selecting the epochs is the one maximizing the overall long-term expected VCC reward. The output of the MDP problem is scheduling solution \textit{ $\pi$ : S $\longrightarrow$ A} that contributes in having a maximum overall long-term rewards that is achieved from the beginning of any state \textit{s}. Our iterative solution calculates the cumulative reward as the output of the function \textit{$V^{*}$}, it characterises the only solution of the Bellman equation as detailed in the Eq. \ref{BellmanEquation}:  
\begin{equation}\label{BellmanEquation}
    V^{*}(s)=\max_{s\in S} [ r(s,a,s') + \sum_{s'\in S}^{}P (s'|s,a)V^{*}(s') ]
\end{equation}
Because of the model is defined as an MDP model with finite  sequential decision to make, a finite set of states and epochs and scheduling actions. Our value iteration algorithm finds the solution by maximizing the problem presented by Eq.\ref{totalReward}. The complete details of the steps run in the algorithm are detailed in Algorithm \textit{3}.
\begin{algorithm}[H]
\begin{algorithmic}[1]
\label{AlgorithmMDPScheme2}
\caption{Algorithm for Value Iteration }
\STATE \textbf{\textit{Step 1:}} We set $V(s)=0$ for every state $s$ of the VCC, $k=0$ and $\epsilon >0$.
\STATE \textbf{\textit{Step2:}} For a state $s$, find $V^{k+1}(s) = \max_{s\in S} [ r(s,a,s') + \sum_{s'\in S}P (s'|s,a)V^{k}(s')]$
\STATE \textbf{\textit{Step 3:}} If $\| V^{k+1}(s) - V^{k}(s) \| < \epsilon$, go to \textbf{\textit{Step 4}}.
Otherwise, add 1 to k and move to \textbf{\textit{Step2}}.
\STATE \textbf{\textit{Step 4:}} For every s $\varepsilon$ S, calculate the optimal scheduling policy for the VCC, stop and output the policy, $\pi$= argmax$_{a\epsilon A_{s}} [ \max_{s\in S} [ r(s,a,s') + \sum_{s'\in S}P (s'|s,a)V^{k+1}(s')]$
\end{algorithmic}
\end{algorithm}
\subsection{Blocks of State Divided Iteration (BSDI) CUDA Accelerated Algorithm of Value Iteration}
GPGPUs are used for highly parallel SIMD compute. The CUDA enabled GPUs guarantees more transistors to be used for data processing in comparison to data flow and caching like in CPU. CUDA Streaming Multiprocessor (SM) have various streaming processors (SPs), or CUDA cores \cite{CUDAByExampleAnIntroductionToGeneralPurposeGPUProgrammingPortableDocuments}, and very efficient in energy \cite{OnThePerformanceAndEnergyEfficiencyOfMultiCoreSIMDCPUsAndCUDAEnabledGPUs}. We us many thread blocks to concurrently execute on a every SM, when done we launch new blocks in free SMs.
In our solution, we focus on the acceleration and the total time to complete the kernel and not the duration it takes a thread to complete. 
\begin{figure}[t]
\begin{center} 
\includegraphics[width=9cm,height=6cm]{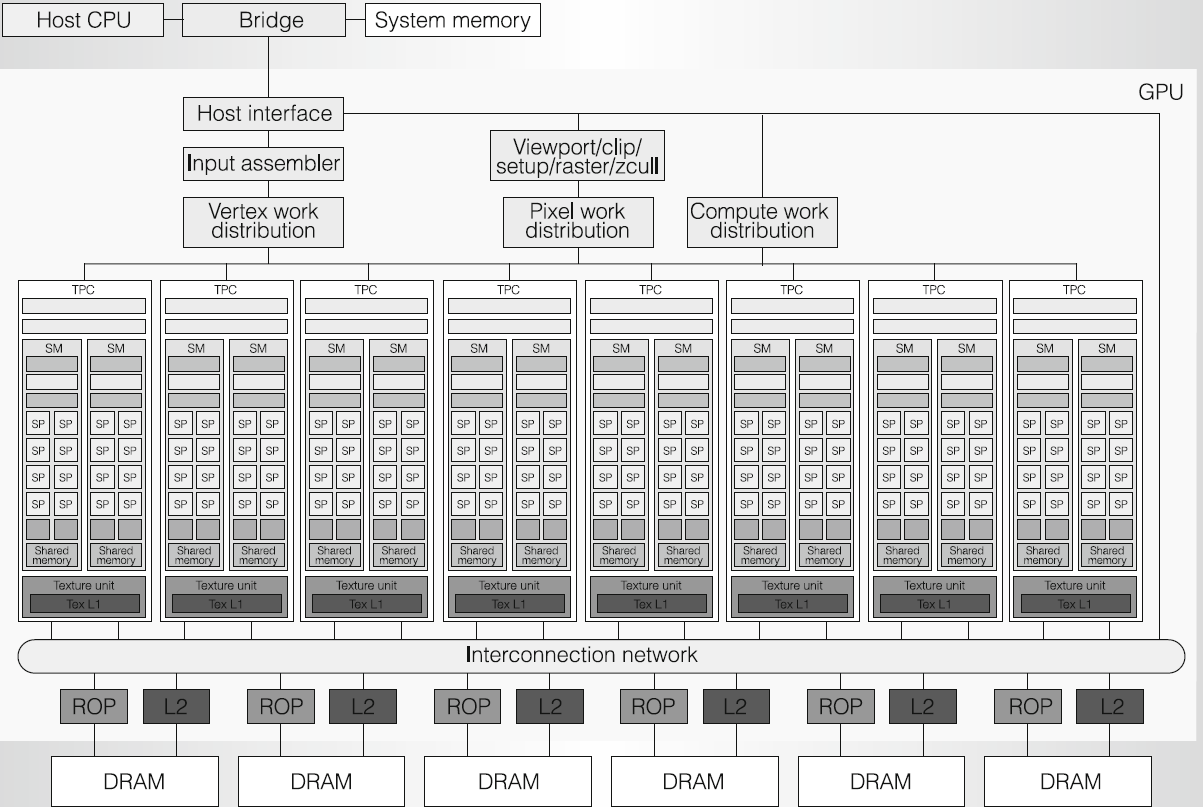}
\end{center}
\caption{Block Diagram of CUDA Enabled GPU.}
\label{LabelWaveChannelArrangement}
\end{figure}
We use NVIDIA's Quadro K1100M GPU, as shown in Table. \ref{TableNvidiaQuadro} to leverage the parallel SIMD capabilities  of CUDA SMs model by running the value iteration algorithm to finding the optimal solution, as initiated in \cite{CUDAAcceleratedTaskSchedulingInVehicularCloudsWithOpportunisticallyAvailableV2I}.
\begin{table}[t]
\begin{center}
\caption{GPU with CUDA characteristics } \label{TableNvidiaQuadro}
\begin{tabular}{ll}
\toprule 
Variable & Description  \\
\midrule
\rowcolor{black!20} 
Device 0/CUDA Driver/Routine & Quadro K1100M with 7.5/6.5 \\
Architecture   & Kepler \\
\rowcolor{black!20} 
CUDA Compute Capability       & 3.0 \\
Total amount of gloal memory  & 2048 MBytes    \\
\rowcolor{black!20} 
Number of MPs       &   2 \\
Number of CUDA cores/MP      & 192     \\
\rowcolor{black!20} 
Total number of cores       & 384 CUDA cores    \\
Clock Rate of GPU       & 706 MHz (0.71 GHz)        \\
\rowcolor{black!20} 
Clock Rate of Memory        & 1400 MHz \\
Memory Bus Width     & 128 bits  \\
\rowcolor{black!20} 
Size of L2 Cache        & 262144 bytes    \\
Total Amount of Constant Memory  & 65536 bytes  \\
\rowcolor{black!20} 
Total Shared Memory per Block  & 49152 bytes    \\
Total Registers per Block  & 65536 \\
\rowcolor{black!20} 
Size of Warp   & 32 \\
Maximum Threads Number per MP  & 2048\\
\rowcolor{black!20} 
Maximum Threads Number per Block  & 1024  \\
Texture alignment & 512 bytes  \\
\rowcolor{black!20} 
Host to Device Bandwidth  & 9735 MB/s  \\
Device to Host Bandwidth & 9302 MB/s  \\
\rowcolor{black!20} 
Device to Device Bandwidth & 3200 MB/s \\
Peak single precision performance & 550 Gigaflops \\
\bottomrule
\end{tabular}
\end{center}
\end{table}
Our BSDI implementation in Algorithm \textit{4} uses parallel set of state space that do not have any relationship to apply the value iteration algorithm in parallel and evaluate the performance of all states after scheduling.
\begin{algorithm}[H]
\caption{Parallel Implementation of BSDI}
\begin{algorithmic}[1]
\label{CudaMdpAcceleration}
    \STATE  cudaMemcpy($V_g(s),h_{V_{g}(s)}$, length$\_$bytes, cudaMemcpyHostToDevice)
    \STATE $V_{g}(s)$ $\gets$  stored state space in global memory 
    \STATE  $\_\_$shared$\_\_$ $V_{s}(s)$ $\gets$  stored state space in shared memory 
    \FORALL{s $\in$ S in parallel}
        \STATE $V_{s}(s)$ = $V_{g}(s)$
        \STATE Synchronize threads : $\_\_$syncthreads()
        \STATE $v$ $\gets$ $V_{s}(s)$
        \FOR{iterations $<$ number of episodes}
            \STATE $\Delta_{s}$ $\gets$ 0 
                \FORALL{a $\in$ A}
                    \STATE $V_{s}(s)\gets \max_{s\in S} [r(s,a,s')+\sum_{s'\in S}^{}P (s'|s,a)V_{InBlock}(s')]$
                \ENDFOR
            \STATE $\Delta_{s}$ $\gets$ $\max{\Delta_{s}, | v  - V_{s}(s)  |}$.
            \STATE $V_{s}(s)$ $\gets$ $v$
            \STATE Synchronize threads : $\_\_$syncthreads()
        \ENDFOR
        \STATE $V_{g}(s)$ = $V_{s}(s)$
    \ENDFOR
        
    \FORALL{ $\Delta_{s}$ of $s \in S$ in parallel}
        \STATE $\pi$ $\gets$ $\max{\Delta_{s}}$   
    \ENDFOR
    
    \STATE cudaMemcpy ($h_\pi$, $\pi$, length$\_$bytes, cudaMemcpyDeviceToHost )
\end{algorithmic}
\end{algorithm}
We first divide the state space \textit{S} in equal sized blocks, then we attribute to every thread in every block to be fulfill the calculus of the reward while transitioning along possible stating after making all the possible scheduling decisions.  
\section{AAA and IEEE 1609.4 simulation and performance comparison under same VCC and the safety services}\label{sec3:SimulationAndComparison}
\begin{figure}[t]
\begin{center}
\includegraphics[width=8cm,height=6cm]{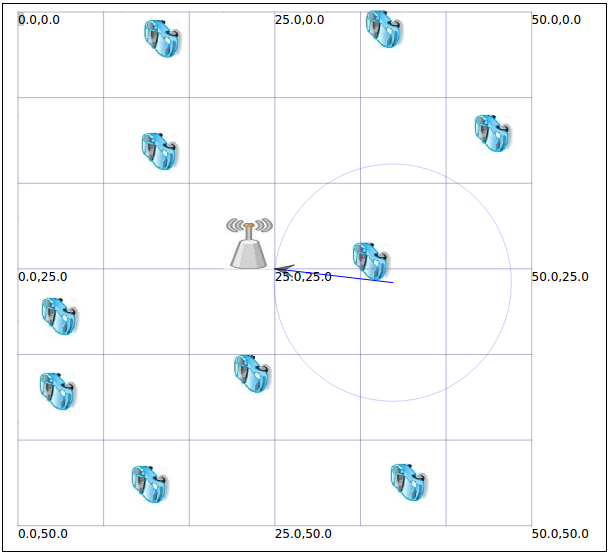}
\caption{NS3 Simulation of the VC Build Around the RSU.}
\label{VCWithRSUgatewaySimulationInNS3}
\end{center}
\end{figure}

In figure \ref{VCWithRSUgatewaySimulationInNS3}, we display screen-shot of the simulation of the VC in NS-3. We create it with an abstract layer for compute of vehicle micro compute and the fixed RSU compute. We randomly distribute the vehicles while moving in way point mobility with variable speeds while kept under RSU coverage. The packets of safety messages are created by vehicles every 100 ms, more precisely it is at the beginning of the first clock tick in the guard intervals before CCHI. Moreover, the packet generation rate of non-safety services in VCs, and are ready for every vehicle at every SCHI. We use for simplicity reason a periodic V2V BSMs that are 200 Bytes in size, same as mandated in one type message in the standard, and we use 1500 Bytes packets of non-safety services. A complete list of parameters that we deployed in the NS3 simulation are described in Table. \ref{ParametersOfTheSimulationInNS3}. For simplicity, we only use same priority for all the packets with the use of a background traffic (BK) like introduced in the access category in EDCA standar for the vehicular safety CCH. 

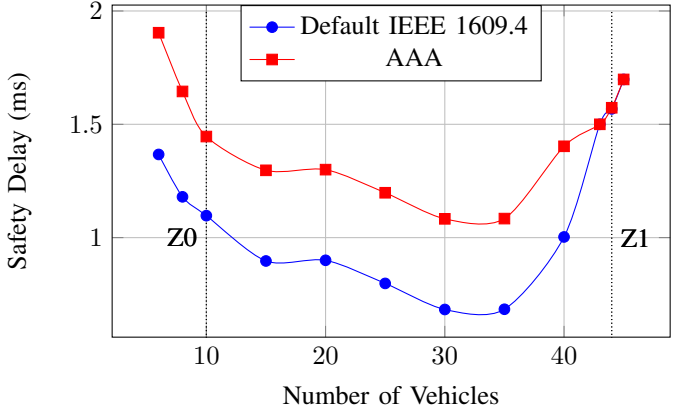
\begin{figure}[t]
\centering
\begin{tikzpicture}
    \pgfplotsset{width=9cm,height=6cm}
    \begin{axis}[
        xlabel=$\text{Number of Vehicles}$,
        ylabel= {Safety Delay (ms) },
        grid=major,
        legend style={at={(0.5,1)},anchor=north}]
    \draw[densely dotted] (axis cs:10,0) -- node[left]{Z0} (axis cs:10,2);
    \draw[densely dotted] (axis cs:10,0) -- node[left]{Z0} (axis cs:10,2);
    \draw[densely dotted] (axis cs:44,0) -- node[right]{Z1} (axis cs:44,2);
    \draw[densely dotted] (axis cs:44,0) -- node[right]{Z1} (axis cs:44,2);
    \addplot[smooth,mark=*,blue] plot coordinates {
    (6,1.367)
    (8,1.18)
    (10,1.097)
    (15,0.897)
    (20,0.9)
    (25,0.798)
    (30,0.683)
    (35,0.684)
    (40,1.003)
    (43,1.5)
    (44,1.568 )
    (45,1.698)
    }; 
    \addlegendentry{Default IEEE 1609.4}
    \addplot[smooth,color=red,mark=square*] plot coordinates {
    (6,1.904)
    (8,1.645)
    (10,1.446)
    (15,1.297)
    (20,1.3)
    (25,1.198)
    (30,1.083)
    (35,1.084)
    (40,1.403)
    (43,1.5)
    (44,1.573 )
    (45,1.698)
    };
    \addlegendentry{AAA}
    \end{axis}
\end{tikzpicture}
\caption{BSMs End-To-End Average Delay Versus Vehicles Count.}%
\label{AverageendtoenddelayofBSMversusnumberofvehicles}
\end{figure}

The Zone Z0 on the left side in Fig. \ref{AverageendtoenddelayofBSMversusnumberofvehicles} has a distant vehicles that are randomly added to the simulation, under the RSU DSRC communication range. Our AAA solution offers a similar average V2V safety packets end-to-end delay to the delay achieved in the default static IEEE 1609.4, even with a reduced CCHI as presented in Fig. \ref{ChannelsIntervalsAdjustmentOfAAAAnd16094VersusTheNumberOfVehicles}. If the number of vehicles introduced within the RSU area, as presented in Z1, we notice the distance between vehicles is reduced and the CCHI is should be increased, as detailed in Fig. \ref{ChannelsIntervalsAdjustmentOfAAAAnd16094VersusTheNumberOfVehicles}.
\begin{table}[t]
    \caption{NS3 SIMULATION PARAMETERS}\label{ParametersOfTheSimulationInNS3}
    \centering
    \begin{tabular}{l l}
        \toprule
        \toprule
        \textbf{Parameters} & \textbf{Values}\\
        Vehicles     & 5 - 50 \\
        Epoch simulation   & 1000 ms          \\         
        RSUs & 1         \\
        VM size  & 500 kbits          \\         
        Synchronisation Interval (SI)  & 100 ms          \\         
        Range of Communication  & 200m \\
        SCHI and CCHI Default   & 50 ms          \\     
        RSU Coverage Area & 50*50m \\
        Default Guard Interval (GI)     & 4 ms         \\         
        Packet Generation Rate for VC & 10 Hz          \\     
        Packet Size for VC  & 1500 Byte          \\     
        Packet rate for BSM    & 10 Hz          \\         
        Packet size for BSM & 200 Byte          \\     
        Data rate and Modulation & QPSK 6 Mbps          \\         
        Bandwidth CCH and SCH & 10 Mhz          \\         
        BK CWmin & 15           \\
        Access Class EDCA  &  Background Traffic (BK) \\
        CWmax of BK & 1023           \\
        Number of SCH and CCH    & 1          \\         
        AIFSN of BK & 9           \\
        Propagation Model     & Nakagami           \\      
\bottomrule
\bottomrule
    \end{tabular}
\end{table}

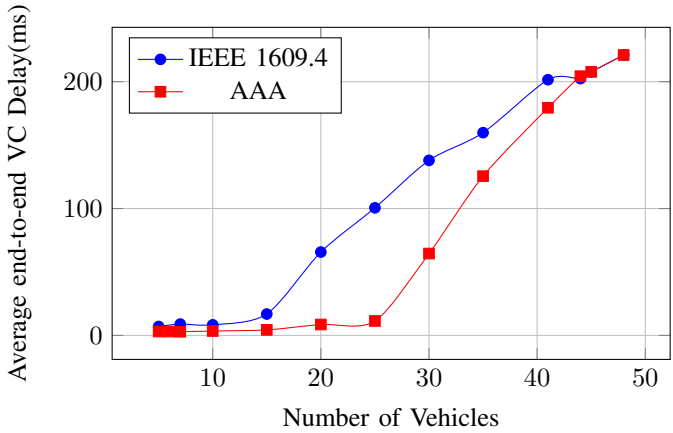
\begin{figure}[h]
\centering
\begin{tikzpicture}
    \pgfplotsset{width=9cm,height=6cm}
    \begin{axis}[
        xlabel=$\text{Number of Vehicles}$,
        ylabel=$\text{Average end-to-end VC Delay(ms)}$,
        grid=major,
        legend pos=north west]  
    \addplot[smooth,mark=*,blue] plot coordinates {
    (5,7.026)
    (7,8.894)
    (10,8.322)
    (15,16.839)
    (20,65.709)
    (25,100.573)
    (30,137.988)
    (35,159.825)
    (41,201.577 )
    (44,202.45 )
    (45,207.762)
    (48,221.068 )
    }; 
    \addlegendentry{IEEE 1609.4}
    \addplot[smooth,color=red,mark=square*] plot coordinates {
    (5,3.043)
    (6,3.043)
    (7,2.883)
    (10,3.438)
    (15,4.359)
    (20,8.59)
    (25,11.356)
    (30,64.531)
    (35,125.539)
    (41,179.463 )
    (44,204.346 )
    (45,207.762)
    (48,221.068 )
    };
    \addlegendentry{AAA}
    \end{axis}
\end{tikzpicture}
\caption{Non-Safety End-To-End Average Delay to RSU.}%
\label{AverageEndToEndDelayToTheRSUOfNonSafetyMessagesVersusNumberOfVehicles}
\end{figure}

\begin{figure}[H]
\centering
\begin{tikzpicture}
    \pgfplotsset{width=9cm,height=6cm}
    \begin{axis}[
        xlabel=$\text{Vehicles Number}$,
        ylabel=$\text{Safety Throughout(Kbps)}$,  
        grid=major,
        legend pos=north west]  
    \addplot[smooth,mark=*,blue] plot coordinates {
    (5,78.4)
    (7,110.4)
    (10,158.4)
    (15,230.4)
    (20,310.4)
    (25,382.4)
    (30,444.8)
    (35,520)
    (40,577.6)
    (42,612.8)
    (43,620.8)
    (45,659.2)
    }; 
    \addlegendentry{Default IEEE 1609.4}
    \addplot[smooth,color=red,mark=square*]
        plot coordinates {
    (5,62.4)
    (7,91.2)
    (10,140.8)
    (15,204.8)
    (20,268.8)
    (25,305.6)
    (30,363.2)
    (35,452.8)
    (40,561.6)
    (42,598.4)
    (43,619.2)
    (45,654.4)
    };
    \addlegendentry{AAA}
    \end{axis}
\end{tikzpicture}
\caption{V2V BSM Safety Packets Throughput Versus Vehicles Count.}%
\label{ThroughputOfV2VBSMSafetyMessagesVersusTheNumberOfVehicles}
\end{figure}
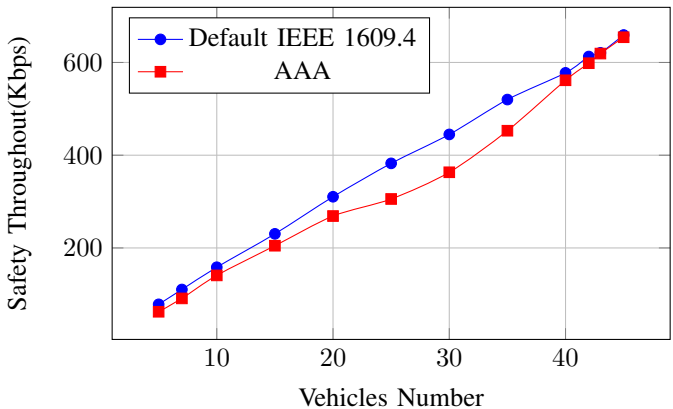

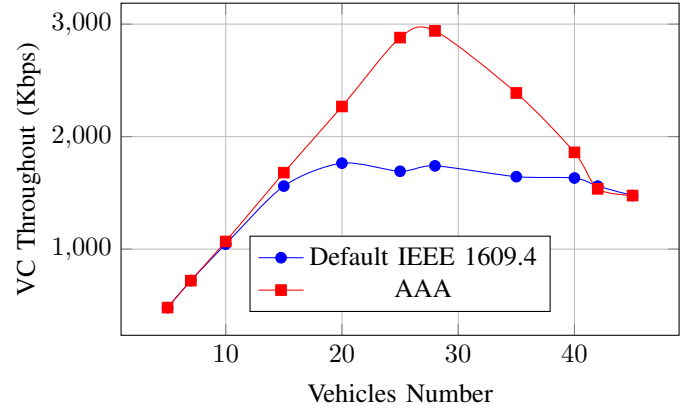
\begin{figure}[H]
\centering
\begin{tikzpicture}
    \pgfplotsset{width=9cm,height=6cm}
    \begin{axis}[
        xlabel=$\text{Vehicles Number}$,
        ylabel=$\text{VC Throughout (Kbps)}$, 
        grid=major,
        legend style={at={(0.5,0.3)},anchor=north}]
    \addplot[smooth,mark=*,blue] plot coordinates {
    (5,480)
    (7,720)
    (10,1044)
    (15,1560)
    (20,1764)
    (25,1692)
    (28,1740)
    (35,1644)
    (40,1632)
    (42,1560)
    (45,1476)
    }; 
    \addlegendentry{Default IEEE 1609.4}
    \addplot[smooth,color=red,mark=square*] plot coordinates {
    (5,480)
    (7,720)
    (10,1068)
    (15,1680)
    (20,2268)
    (25,2880)
    (28,2940)
    (35,2388)
    (40,1860)
    (42,1536)
    (45,1476)
    };
    \addlegendentry{AAA}
    \end{axis}
\end{tikzpicture}
\caption{Throughput of VC Versus Vehicles Number.}\label{VcThroughputOfNonSafetyApplications}
\label{VCThroughputOfNonSafetyApplicationsVersusNumberOfVehicles}
\end{figure}

The average end-to-end delay of packets delivered in the VC is acceptable in low dense VANETs, as shown in Fig. \ref{AverageEndToEndDelayToTheRSUOfNonSafetyMessagesVersusNumberOfVehicles}. When the number of vehicles gets higher, we notice more BSMs are queued and the CCHI has to get larger so that we lower the number of unsent queued packets. Consequently, this reduces the II and decreases the transmission during SCHI for VC packets, as detailed in Fig. \ref{ChannelsIntervalsAdjustmentOfAAAAnd16094VersusTheNumberOfVehicles}, and will increase the VC packets average end-to-end delay.
Our proposed AAA Multi-Channel operations solutions has a comparable V2V throughput and successful BSMs exchange rate as achieved in IEEE 1609.4 standar with static access switching, a detailed comparison is presented in Fig. \ref{ThroughputOfV2VBSMSafetyMessagesVersusTheNumberOfVehicles}, while CCHI was decreased significantly as shown in Fig. \ref{ChannelsIntervalsAdjustmentOfAAAAnd16094VersusTheNumberOfVehicles}. 
We achieved a significant better throughput for VC using our AAA scheme over the static standard IEEE 1609.4 as detailed in Fig. \ref{VCThroughputOfNonSafetyApplicationsVersusNumberOfVehicles}. The better achieved results are related to dynamically changing the CCHI by the detected II, it is significant when the numbers of vehicles is low in VANETs. Hence, once we extend the SCHI and increase the SCH access for VC packets transmission. 

Fig. \ref{ChannelsIntervalsAdjustmentOfAAAAnd16094VersusTheNumberOfVehicles} shows the impact of the AAA scheme on finding the inactivity interval and automatically adapting the interval timers of IEEE 1609.4 by decreasing the CCHI for BSM dissemination and increasing the SCHI for VC usage for both types of vehicular communication in WAVE.
The formulation of the average throughput per vehicle of non-safety applications that are hosted in the VC is given by Eq.\ref{AvgThrouPerVehicle} as follows : 
\begin{equation} \label{AvgThrouPerVehicle}
Th_{ij}= \frac{Th_{VCij}}{j} 
\end{equation}
where $Th_{VCij}$ represents the calculated throughput of the VC found while using the $i^{th}$ mode of either the default IEEE 1609.4 or the AAA scheme for a number of \textit{j} OBUs forming it. 
\begin{figure}[H]
\centering
\begin{tikzpicture}
    \pgfplotsset{width=9cm,height=6cm}
    \begin{axis}[
        xlabel=$\text{Number of Vehicles}$,
        ylabel=$\text{Interval Time}$, 
        grid=major,
        legend pos=north east]  
    \addplot[smooth,mark=*,blue] plot coordinates {
    (5,50)
    (7,50)
    (10,50)
    (15,50)
    (20,50)
    (25,50)
    (30,50)
    (35,50)
    (40,50)
    (42,50)
    (43,50)
    (45,50)
    }; 
    \addlegendentry{Default IEEE 1609.4}
    \addplot[smooth,color=red,mark=square*] plot coordinates {
    (5,9.53)
    (7,10.57)
    (10,13.06)
    (15,15.35)
    (20,20.4)
    (25,22.369)
    (30,20.16)
    (35,26.30)
    (40,43.34)
    (42,47.19)
    (43,49.08)
    (45,50)
    };
    \addlegendentry{AAA CCHI}
    \addplot[smooth,color=pink,mark=square*] plot coordinates {
    (5,90.46)
    (7,89.42)
    (10,86.93)
    (15,84.64)
    (20,79.59)
    (25,77.59)
    (30,79.83)
    (35,73.69)
    (40,56.65)
    (42,52.80)
    (43,50.91)
    (45,50)
    };
    \addlegendentry{AAA SCHI}
    \end{axis}
\end{tikzpicture}
\caption{Adjusted Channels Intervals for  AAA and IEEE 1609.4 Versus Vehicles Number }
\label{ChannelsIntervalsAdjustmentOfAAAAnd16094VersusTheNumberOfVehicles}
\end{figure}
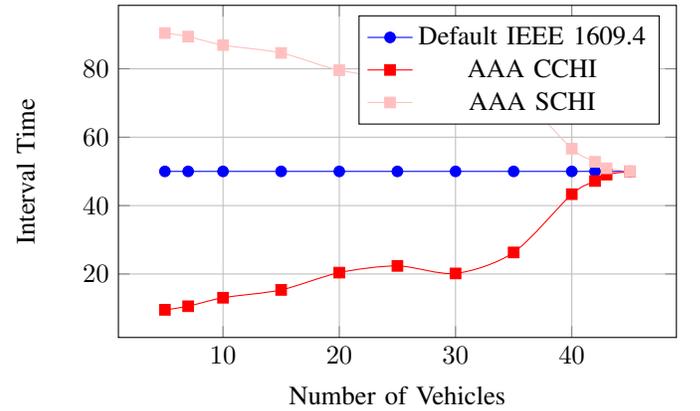
\begin{figure*}[t]
\begin{center}
\begin{tikzpicture}
                \begin{axis}[view={140}{40},
            	width=220pt,
            	height=230pt,
            	grid=major,
            	x dir=reverse,
            	z buffer=sort,
            	enlargelimits=upper,
            	xlabel=$\text{Vehicles}$,
            	ylabel=$\text{Time (s)}$,
            	zlabel=$\text{Avg Dropped VM (\%)}$,
            	xtick={5,10,...,60},
            	ytick={1,3,...,16},
            	ztick={0,10,20,...,100},]
            	\addplot3+[mesh,scatter,samples=10,mark=cube*,domain=0:1] 
            		table {1DefaultOddMltiChannel.dat};
            	\addplot3+[mesh,scatter,samples=10,mark=cube*,domain=0:1] 
            		table {2DefaultOddMltiChannel.dat};
                \addplot3+[mesh,scatter,samples=10,mark=cube*,domain=0:1] 
            		table {3DefaultOddMltiChannel.dat};
            	\addplot3+[mesh,scatter,samples=10,mark=cube*,domain=0:1] 
            		table {4DefaultOddMltiChannel.dat};
                \addplot3+[mesh,scatter,samples=10,mark=*,domain=0:1] 
            		table {1DefaultevenMltiChannel.dat};
                \addplot3+[mesh,scatter,samples=10,mark=*,domain=0:1] 
                		table {2DefaultevenMltiChannel.dat};
                \addplot3+[mesh,scatter,samples=10,mark=*,domain=0:1] 
                		table {3DefaultevenMltiChannel.dat};
                \addplot3+[mesh,scatter,samples=10,mark=*,domain=0:1] 
                		table {4DefaultevenMltiChannel.dat};
                \addplot3+[mesh,scatter,samples=10,mark=*,domain=0:1] 
                		table {5DefaultevenMltiChannel.dat};
                \addlegendentry{Multi-Channel Operations IEEE 1609.4 Standard}
            	\end{axis}
\end{tikzpicture}%
\begin{tikzpicture}
                \begin{axis}[view={140}{40},
            	width=220pt,
            	height=230pt,
            	grid=major,
            	x dir=reverse,
            	z buffer=sort,
            	enlargelimits=upper,
            	xlabel=$\text{Vehicles}$,
            	ylabel=$\text{Time (s)}$,
            	zlabel=$\text{Avg Dropped VM (\%)}$,
            	xtick={5,10,...,60},
            	ytick={1,3,...,16},
            	ztick={0,10,20,...,100},]
            	\addplot3+[mesh,scatter,samples=10,mark=cube*,domain=0:1] 
            		table {1AAAOddMltiChannel.dat};
            	\addplot3+[mesh,scatter,samples=10,mark=cube*,domain=0:1] 
            		table {2AAAOddMltiChannel.dat};
                \addplot3+[mesh,scatter,samples=10,mark=cube*,domain=0:1] 
            		table {3AAAOddMltiChannel.dat};
            	\addplot3+[mesh,scatter,samples=10,mark=cube*,domain=0:1] 
            		table {4AAAOddMltiChannel.dat};
                \addplot3+[mesh,scatter,samples=10,mark=*,domain=0:1] 
            		table {1AAAevenMltiChannel.dat};
                \addplot3+[mesh,scatter,samples=10,mark=*,domain=0:1] 
                		table {2AAAevenMltiChannel.dat};
                \addplot3+[mesh,scatter,samples=10,mark=*,domain=0:1] 
                		table {3AAAevenMltiChannel.dat};
                \addplot3+[mesh,scatter,samples=10,mark=*,domain=0:1] 
                		table {4AAAevenMltiChannel.dat};
                \addplot3+[mesh,scatter,samples=10,mark=*,domain=0:1] 
                		table {5AAAevenMltiChannel.dat};
                \addlegendentry{AAA Multi-Channel Operations}
            	\end{axis}
\end{tikzpicture}
\end{center}
\caption{Average Percentage of Dropped VMs.}
\label{AveragePercentageOfDroppedVMsInDifferentNetwrokDensityVersusTime}
\end{figure*}
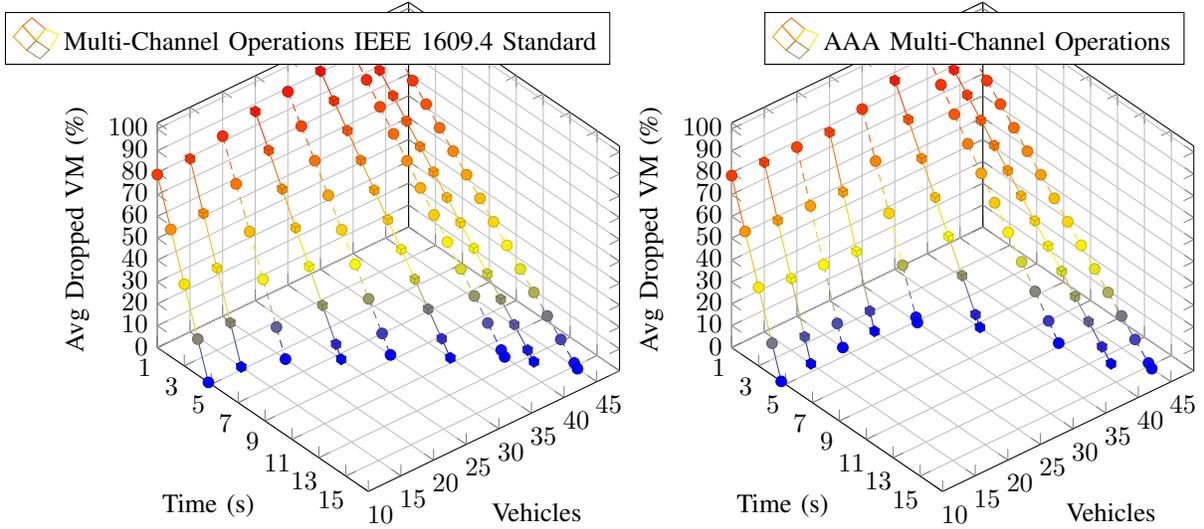
The migration scenario consists of a vehicle that has a total of $T_{VMsize}$ = 500 kbits as the size of all its VMs running in its OBU and should do migration to the RSU before leaving the network and the VC. 
The average percentage of dropped VMs, are presented in Fig. \ref{AveragePercentageOfDroppedVMsInDifferentNetwrokDensityVersusTime}, at an instant t(s) from the beginning of migration is defined as shown in Eq. \ref{AvgDropedVM}.   \begin{equation} \label{AvgDropedVM}
AVM_{t}=  1- \sum_{k=0}^{t}\frac{  T_{VMsize} - k * Th_{VCij}} {Th_{VCij}}  
\end{equation}
\section{Numerical Results and Analysis of the scheduling schemes in VCC} \label{NumericalResultsAndAnalysismdpgreedySIMULATION}
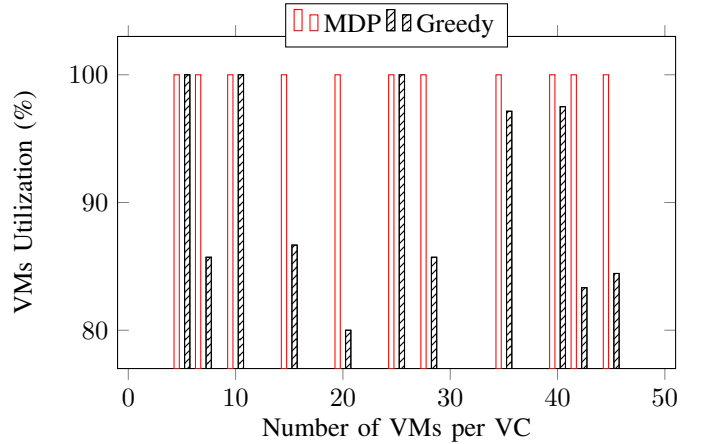
\begin{figure}[t]
\centering
\begin{tikzpicture}
\pgfplotsset{width=9cm,height=6cm}
\begin{axis}[
	x tick label style={
		/pgf/number format/1000 sep=},
	ylabel=VMs Utilization (\%),
	xlabel=Number of VMs per VC ,
	enlargelimits=0.15,
	legend style={at={(0.5,1.1)},
		anchor=north,legend columns=-1},
	ybar,
	bar width=2pt,
]
\addplot[color=red] coordinates {
    (5,100)
    (7,100)
    (10,100)
    (15,100)
    (20,100)
    (25,100)
    (28,100)
    (35,100)
    (40,100)
    (42,100)
    (45,100)
};
\addplot [postaction={pattern=north east lines}]
	coordinates {
    (5,100)
    (7,85.7143)
    (10,100)
    (15,86.6667)
    (20,80)
    (25,100)
    (28,85.7143)
    (35,97.1429)
    (40,97.5)
    (42,83.3333)
    (45,84.4444)
};
\legend{MDP,Greedy}
\end{axis}
\end{tikzpicture}
\caption{VCC VMs Utilization for Greedy and MDP Schemes.}%
\label{VCCVMsUtilizationForGreedyAndMDPSchemes}
\end{figure}

We evaluate the scheduling schemes for different VCCs and the scheduling epochs made, based on the MDP and greedy heuristic, and we consider and compare the resource usage of VMs in the VCC and the overall calculate reward after all the scheduling decisions of the task is over.
In Fig. \ref{VCCVMsUtilizationForGreedyAndMDPSchemes}, we show the histogram detailing the total utilization percentage of VMs after tasks scheduling from the BOTs while using the MDP-based and the fast greedy schemes. We note that having equal VM utilization in VCs from the VCC under the greedy scheme and MDP scheduling does not necessarily reflect similar placement decisions. That is, many of the VCs forming the VCC system contain left behind/unused VMs under the greedy algorithm since it does not calculate the impact of every decision on the next placement decisions. The VM utilization achieved by using the MDP scheduling scheme is maximized through the selection of the highest total reward of a sequence of placement decisions, rather than the reward per task scheduling.     
\newline
\newline

Based on the VCC utilization and the scheduling decisions made by the greedy algorithm, Fig. \ref{VCThroughputOfNonSafetyApplicationsVersusNumberOfVehicles} shows that soms VCs did not reach their optimal throughput under both of the IEEE 1609.4 and the AAA schemes. This is expected since the heuristic does not guarantee optimality.
\begin{figure}[t]
\centering
\begin{tikzpicture}
\pgfplotsset{width=9cm,height=6cm}
    \begin{axis}[stack plots=y,/tikz/ybar, bar width=0.2pt, yticklabels={,,}, xticklabels={,,}]
    	\addplot coordinates {(5,0) (7,0)   (10,0)   (15,0)   (20,0)     (25,0)    (28,0)     (35,0)      (40,0)      (42,0)       (45,0)         };
    	\addplot coordinates {(5,0) (7,174) (10,213) (15,288) (20,362.8) (25,576)  (28,535.5) (35,397.08) (40,315.7)  (42,242.8)   (45,247.5)     };
    	\addplot coordinates {(5,0) (7,1) (10,213) (15,288) (20,362.8) (25,576)  (28,535.5) (35,463.26) (40,360.8)  (42,273.15)  (45,247.5)     };
    	\addplot coordinates {(5,0) (7,1) (10,320) (15,288) (20,362.8) (25,576)  (28,535.5) (35,463.26) (40,360.8)  (42,273.15)  (45,275)       };
    	\addplot coordinates {(5,0) (7,0)   (10,0) (15,100) (20,462.8) (25,576)  (28,735.5) (35,463.26) (40,260.8)  (42,273.15)  (45,175)       };
    	\addplot coordinates {(5,0) (7,0)   (10,0)   (15,0)   (20,0)     (25,576)  (28,0)     (35,363.26) (40,360.8)  (42,0)       (45,0)         };
      \end{axis}
    \begin{axis}[
        xlabel=$\text{Number of VMs per VC}$,
        grid=major,
        ylabel=$\text{VC Throughout (Kbps)}$, 
    	legend style={at={(0.5,1.1)},
		anchor=north,legend columns=-1}]
    \addplot[smooth,color=red,mark=square*] plot coordinates {
    (5,480)
    (7,612)
    (10,1068)
    (15,1444.8) 
    (20,1814.4) 
    (25,2880)   
    (28,2499)   
    (35,2316.36)
    (40,1804.2) 
    (42,1274.88)
    (45,1239.84)
    };
    \addlegendentry{AAA Scheme}
    \addplot[smooth,mark=*,blue] plot coordinates {
    (5,480)      
    (7,612)      
    (10,1044)    
    (15,1341.6)  
    (20,1411.2)  
    (25,1692)    
    (28,1479)    
    (35,1594.68) 
    (40,1583.04) 
    (42,1294.8)  
    (45,1239.84) 
    }; 
    \addlegendentry{Default 1609.4}
    \end{axis}
\end{tikzpicture}
\centering
\caption{Vehicular Cloud Throughput Under Greedy Scheduling.}
\label{VCThroughputOfNonSafetyApplicationsVersusNumberOfVehicles}
\end{figure}
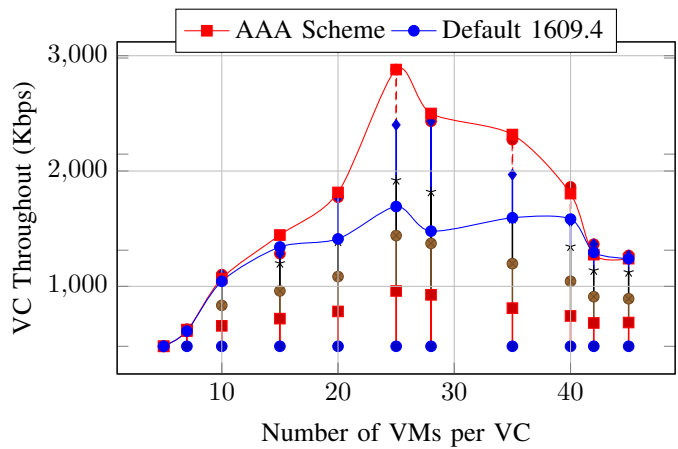

\begin{figure}[t]
\centering
\begin{tikzpicture}
    \pgfplotsset{width=9cm,height=6cm}
    \begin{axis}[
        xlabel=$\text{Number of Vehicles}$,
        ylabel=$\text{VC Throughout (Kbps)}$, 
        grid=major,
    	legend style={at={(0.5,1.1)},
		anchor=north,legend columns=-1}]
    \addplot[smooth,mark=*,blue] plot coordinates {
    (5,480)
    (7,720)
    (10,1044)
    (15,1560)
    (20,1764)
    (25,1692)
    (28,1740)
    (35,1644)
    (40,1632)
    (42,1560)
    (45,1476)
    }; 
    \addlegendentry{Default 1609.4}
    \addplot[smooth,color=red,mark=square*] plot coordinates {
    (5,480)   
    (7,720)   
    (10,1068) 
    (15,1680) 
    (20,2268) 
    (25,2880) 
    (28,2940) 
    (35,2388) 
    (40,1860) 
    (42,1536) 
    (45,1476) 
    };
    \addlegendentry{AAA}
    \end{axis}
    \begin{axis}[legend style={at={(0.5,1.1)},anchor=north,legend columns=-1},
        stack plots=y,/tikz/ybar, bar width=0.2pt, yticklabels={,,}, xticklabels={,,}]
    	\addplot coordinates {(5,0)  (7,0)     (10,0)     (15,0)   (20,0)     (25,0)      (28,0)   (35,0)      (40,0)      (42,0)       (45,0)     };
    	\addplot coordinates {(5,96) (7,205.7) (10,427.2) (15,224) (20,680.4) (25,345.6)  (28,420) (35,545.76) (40,93)     (42,36.57)   (45,32.8)     };
    	\addplot coordinates {(5,0)  (7,0)     (10,240.8) (15,672) (20,793.8) (25,806.4)  (28,840) (35,613.98) (40,418.5)  (42,365.7)   (45,360.8)    };
    	\addplot coordinates {(5,0)  (7,0)     (10,0)     (15,584) (20,693.8) (25,806.4)  (28,840) (35,613.98) (40,418.5)  (42,365.7)   (45,360.8)    };
    	\addplot coordinates {(5,0)  (7,0)     (10,0)     (15,0)   (20,0)     (25,921.6)  (28,840) (35,450.98) (40,465)    (42,365.7)   (45,360.8)    };
    	\addplot coordinates {(5,0)  (7,0)     (10,0)     (15,0)   (20,0)     (25,0)      (28,0)   (35,0)      (40,265)    (42,150.27)  (45,100.8)    };
    \legend{Default 1609.4, AAA}
    \end{axis}
    
\end{tikzpicture}
\caption{VC Throughput Under MDP Scheduling.}%
\label{VehicularCloudBasedApplicationThroughputUnderMDPScheduling}
\end{figure}
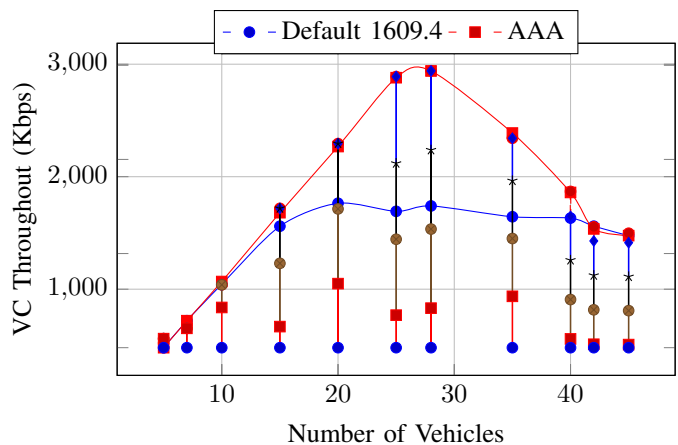

The evaluation of VCs' throughput by using the MDP scheme presents, as described in Fig. \ref{VehicularCloudBasedApplicationThroughputUnderMDPScheduling}, is a maximized effective utilization of the VMs. Moreover, the tasks hosted in the VMs do not follow a specific rule as in the greedy scheme and highly depend on decisions with the highest total reward of the cloud (i.e. fill up the maximum available VMs in every VC).

\begin{figure}[t]
\centering
    \begin{tikzpicture}
    \pgfplotsset{width=9cm,height=7cm}
    \begin{axis}[
        grid=major,ybar stacked,enlargelimits=0.15,
        legend style={at={(0.5,+1.1)},font=\tiny,anchor=north,legend columns=-1},
        ylabel={Virtual Machines}, symbolic x coords={1, 2, 3, 4, 5, 6, 7, 8, 9, 10, 11, , VCC, TCC},
        xtick=data, x tick label style={rotate=45,anchor=east}, ytick={0,20,...,100},ymin=0,ymax=100,xlabel={Tasks in BOTs}]
    \addplot+[ybar,postaction={pattern=north east lines}] plot coordinates { (1,5) (2,10) (3,15) (4,20) (5,25) (6,30) (7,35) (8,40) (9,45) (10,20) (11,0) (VCC, 0) (TCC, 0) };
    \addplot+[ybar,postaction={pattern=dots}] plot coordinates { (1,0) (2,0)  (3,0)  (4,0)  (5,0)  (6,0)  (7,0)  (8,0)  (9,0) (10,0) (11,0) (VCC, 0) (TCC, 85) };
    \addplot+[ybar,postaction={pattern=grid}] plot coordinates { (1,0) (2,0)  (3,0)  (4,0)  (5,0)  (6,0)  (7,0)  (8,0)  (9,0)  (10,0) (11,0) (VCC, 27) (TCC, 0) };
    \addplot+[ybar,postaction={pattern=horizontal lines}] plot coordinates { (1,0) (2,0)  (3,0)  (4,0)  (5,0)  (6,0)  (7,0)  (8,0)  (9,0)  (10,30) (11,55) (VCC, 0) (TCC, 0)};
    \legend{VC VMs, Tradional VMs , Underutilized VMs, Paid VMs}
    \end{axis}
    \end{tikzpicture}
\centering
\caption{Scheduling BOTs with Greedy Solution.}
\label{BOTsPlacementsUnderGreedyScheme}
\end{figure}
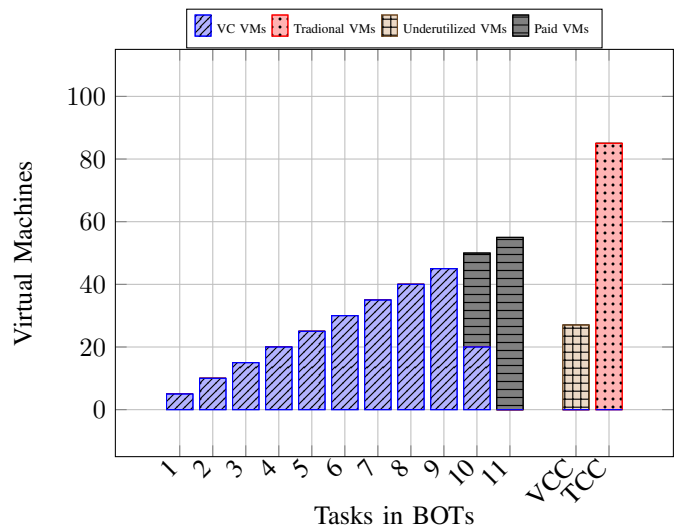

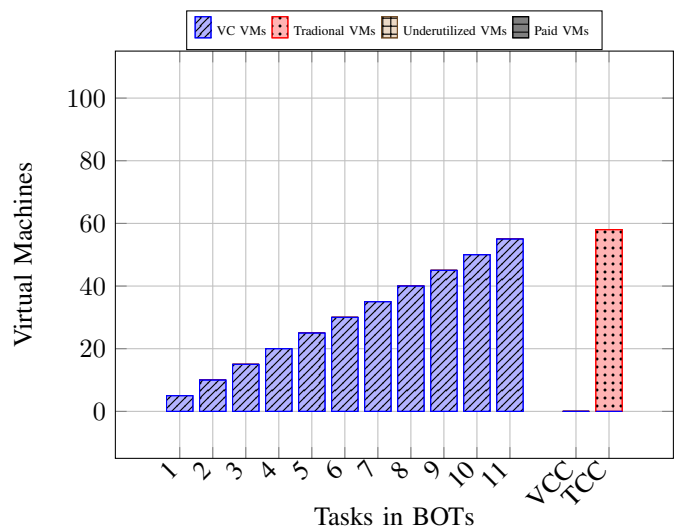
\begin{figure}[t]
\centering
    \begin{tikzpicture}
    \pgfplotsset{width=9cm,height=7cm}
    \begin{axis}[
        grid=major,ybar stacked,enlargelimits=0.15,
        legend style={at={(0.5,+1.1)},font=\tiny,anchor=north,legend columns=-1},
        ylabel={Virtual Machines}, symbolic x coords={1, 2, 3, 4, 5, 6, 7, 8, 9, 10, 11, , VCC, TCC},
        xtick=data, x tick label style={rotate=45,anchor=east},ytick={0,20,...,100},ymin=0,ymax=100, xlabel={Tasks in BOTs}]
    \addplot+[ybar,postaction={pattern=north east lines}] plot coordinates { (1,5) (2,10) (3,15) (4,20) (5,25) (6,30) (7,35) (8,40) (9,45) (10,50) (11,55) (VCC, 0) (TCC, 0) };
    \addplot+[ybar,postaction={pattern=dots}] plot coordinates { (1,0) (2,0)  (3,0)  (4,0)  (5,0)  (6,0)  (7,0)  (8,0)  (9,0) (10,0) (11,0) (VCC, 0) (TCC, 58) };
    \addplot+[ybar,postaction={pattern=grid}] plot coordinates { (1,0) (2,0)  (3,0)  (4,0)  (5,0)  (6,0)  (7,0)  (8,0)  (9,0)  (10,0) (11,0) (VCC, 0) (TCC, 0) };
    \addplot+[ybar,postaction={pattern=horizontal lines}] plot coordinates { (1,0) (2,0)  (3,0)  (4,0)  (5,0)  (6,0)  (7,0)  (8,0)  (9,0)  (10,0) (11,0) (VCC, 0) (TCC, 0) };
    \legend{VC VMs, Tradional VMs , Underutilized VMs, Paid VMs}
    \end{axis}
    \end{tikzpicture}
\centering
\caption{Scheduling BOTs With MDP-Based Solution.}%
\label{BOTsschedulingUnderMDPBasedScheme}
\end{figure}

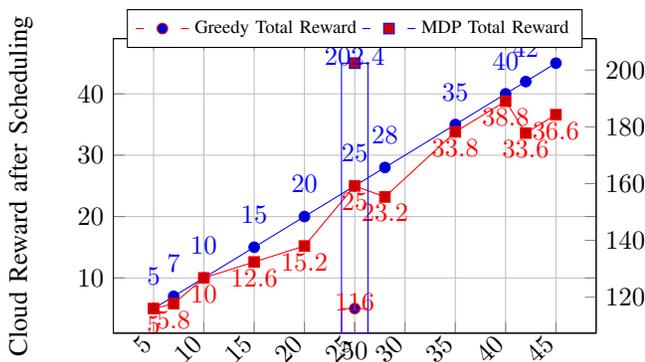
\begin{figure}[t]
    \begin{tikzpicture}
        \centering
        \pgfplotsset{width=8.cm,height=5.5cm}
        \begin{axis} [ylabel=Cloud Reward after Scheduling ,
        legend style={at={(0.5,+1.1)},anchor=north,legend columns=-1},
        ytick={0,10,20,30,40,50},xmajorgrids,ymajorgrids,x tick label style={rotate=45,anchor=east},
        xtick={5, 10, 15, 20, 25, 30, 35, 40, 45, 50 }]
    	\addplot+[draw, nodes near coords, every node near coord/.append style={yshift=+0.2cm}] coordinates {
    	(5,5) (7,7) (10,10) (15,15) (20,20) (25,25) (28,28) (35,35) (40,40) (42,42) (45,45)};
        \addplot+[draw,nodes near coords, every node near coord/.append style={yshift=-0.45cm}] coordinates {
    	(5,5) (7,5.8) (10,10) (15,12.6) (20,15.2) (25,25) (28,23.2) (35,33.8) (40,38.8) (42,33.6) (45,36.6) };
        \end{axis}
        \begin{axis}[axis y line*=right,legend style={at={(0.5,-0.10)},anchor=north},
        legend style={at={(0.5,+1.1)},font=\scriptsize,anchor=north,legend columns=-1},
        xtick=50] 
            \addplot+[ybar,color=red,nodes near coords, every node near coord/.append style={yshift=-0.15cm}]  plot coordinates{(50, 116)};
            \addplot+[ybar,color=blue,nodes near coords, every node near coord/.append style={yshift=-0.15cm}] plot coordinates {(50, 202.4)};
        \legend{Greedy Total Reward, MDP Total Reward}
        \end{axis}
    \end{tikzpicture}
    \centering
    \caption{ VC Reward and Overall VCC.}
    \label{VCsandoverallVCCRewardMDPandGreedy}
\end{figure}

\begin{figure}[t]
\centering
\begin{tikzpicture}
    \pgfplotsset{width=8.cm,height=5.5cm}
    \begin{axis}[
        axis y line*=left,
        axis x line=bottom,
        major x tick style = transparent,
        ybar=5*\pgflinewidth,
        bar width=14pt,
        ymajorgrids = true,
        ylabel = {},
        xlabel = {},
        symbolic x coords={VIA,BSID},
        xtick = data,
            scaled y ticks = false,
            enlarge x limits=0.25,
            axis line style={-},
            ymin=-0.1,ymax=160,
        legend columns=2,
        legend cell align=left,
        legend style={
                at={(0.5,-0.2)},
                anchor=north,
                column sep=1ex
        }
    ]
        \addplot[style={blue,fill=blue,mark=none}]
           coordinates {(VIA, 156.8 ) (BSID, 0)  };
        \addplot[style={red,fill=red,mark=none}]
             coordinates {(VIA, 0) (BSID, 106.08)};
        \legend{Sequential, CUDA-accelerated}
    \end{axis}
    \begin{axis}[
        axis y line*=right,
        axis x line=bottom,
        major x tick style = transparent,
        ybar=5*\pgflinewidth,
        bar width=14pt,
        ylabel = {Total Convergence Time(ms)},
        xlabel = {},
        symbolic x coords={Speedup},
        xtick = data,
            scaled y ticks = false,
            enlarge x limits=0.25,
            axis line style={-},
            ymin=-0.1,ymax=100,
        legend columns=2,
        legend cell align=left,
        legend style={
                at={(0.5,-0.2)},
                anchor=north,
                column sep=1ex
        }
    ]
    \addplot[style={green,fill=green,mark=none}]
             coordinates {(Speedup,32)};
    \end{axis}
\end{tikzpicture}
    \caption{CUDA-Accelerated and Sequential Speedup and Convergence Time.}
    \label{SpeedupAndConvergenceTime}
\end{figure}
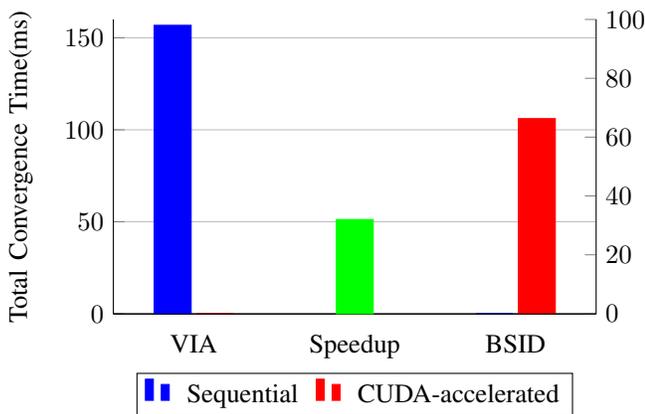

Fig. \ref{BOTsPlacementsUnderGreedyScheme} and Fig. \ref{BOTsschedulingUnderMDPBasedScheme} show how the tasks of the BOTs are distributed along free VMs in VCs and paid VMs in TCC. In addition, we compare the total underutilized VMs left by the greedy and MDP-based scheme in the VCC and total required VMs for both schemes. In the same cloud environments, we present the achieved improvements in terms of the number of occupied VMs that MDP model offers in scheduling over the fast greedy scheme,  
The maximized reward with optimal task scheduling decisions in all BOTs under the greedy heuristic and MDP schemes is detailed in Fig. \ref{VCsandoverallVCCRewardMDPandGreedy}. We present a bar chart with the overall reward of the entire VCC with cost of under-utilization of VMs included. We present a dotted chart to present the reward per VC. We compare our developed MDP scheduling decisions outperforming the heuristic greedy solution for not only most VCs but also in the total gained reward of the entire VCC. Both for utilization and reward, our solution is better and comes with an improved revenue and lower cost of scheduling decisions $\beta_{vc}$=1, $\beta_{tc}$=1.2 and $\gamma_{vc}$=1.
The implemented BSDI solution has 124 threads per block and 124/192=64.5\% occupancy per MP. We present in Fig. \ref{SpeedupAndConvergenceTime}, that the parallel implementation of BSID performs faster and retrieve results quicker than the sequential implementation while finding the same optimal scheduling epochs. Compute performance factor is important when finding solution to MDPs especially that size of explored states can increase tremendously.
The developed BSID implementation has a 32$\%$ speedup compared to the sequential one while reaching same optimal tasks scheduling decisions. The achieved speedup can be improved if we use a better GPU with a higher number of SM.

\section{Conclusion}\label{SectionConclusion}
The emerging concept of Internet of Vehicles (IoV) is encountering many technological challenges to meet the reliability of safety communication between vehicles and the performance of the vehicular cloud computing (VCC) for a better driving experience. In this paper, we developed a  static switching mechanism that considers different types of vehicular applications to improve scheduling. We studied a realistic problem of bag-of-tasks (BOTs) scheduling with different VM requirements onto opportunistically available V2I for vehicular clouds (VCs). We presented a parallel value iteration algorithm using CUDA-enabled GPUs to find optimal scheduling decisions as a solution of the MDP scheme, which outperforms greedy heuristic placements in both of the vehicular cloud resource utilization and the reward. This work can be further extended to study efficient task placement with consideration of real-time delay-intolerant application requirements in the distributed VC forming a VCC.

\bibliographystyle{IEEEtran}
\bibliography{main}
\end{document}